# Multimodal Learning for Just-In-Time Software Defect Prediction in Autonomous Driving Systems

Faisal Mohammad
Department of Software Engineering
Jeonbuk National University
Jeonju, South Korea
mfaisal@jbnu.ac.kr

Duksan Ryu
Department of Software Engineering
Jeonbuk National University
Jeonju, South Korea
duksan.ryu@jbnu.ac.kr

*Abstract*— **In recent years, the rise of autonomous driving technologies has highlighted the critical importance of reliable software for ensuring safety and performance. This paper proposes a novel approach for just-in-time software defect prediction (JIT-SDP) in autonomous driving software systems using multimodal learning. The proposed model leverages the multimodal transformers in which the pre-trained transformers and a combining module deal with the multiple data modalities of the software system datasets such as code features, change metrics, and contextual information. The key point for adapting multimodal learning is to utilize the attention mechanism between the different data modalities such as text, numerical, and categorical. In the combining module, the output of a transformer model on text data and tabular features containing categorical and numerical data are combined to produce the predictions using the fully connected layers. Experiments conducted on three open-source autonomous driving system software projects collected from the GitHub repository (Apollo, Carla, and Donkeycar) demonstrate that the proposed approach significantly outperforms state-of-the-art deep learning and machine learning models regarding evaluation metrics. Our findings highlight the potential of multimodal learning to enhance the reliability and safety of autonomous driving software through improved defect prediction.**

## I. INTRODUCTION

Recent advances in the efficacy of artificial intelligence (AI), the adoption of IoT devices and the power of edge computing have come together to unlock the power of Edge-AI. Edge computing devices equipped with the AI technology provide revolutionary benefits with high-performance computing and a scalable infrastructure. The rapid advancement of autonomous driving systems (ADS) has brought about a paradigm shift in the automotive industry, promising enhanced safety, efficiency, and convenience. However, the complexity of AI-enabled ADS software, often comprising millions of lines of code, poses significant challenges for ensuring its reliability and dependability. Software defects in ADS can lead to catastrophic consequences, including accidents, injuries, and even fatalities. Therefore, effective software defect prediction is paramount to mitigate risks and ensure the safe deployment of ADS. In the transportation domain, leading tech companies like Nvidia, Baidu, Google, etc. have created a platform which integrates edge cloud systems and AI-based in-vehicle hardware and software to deliver scalable and robust autonomy [1].

Just-in-Time (JIT) software defect prediction is an essential technique in software engineering that aims to identify defective code segments as they are introduced into the system, rather than waiting until later stages of development. In AI-enabled systems, such as autonomous driving systems, JIT defect prediction plays a critical role due to the need for real-time, high-assurance software functionality. Autonomous systems operate in safety-critical environments where software defects can have severe consequences, such as malfunctions in vehicle control systems, leading to potentially life-threatening scenarios. Therefore, ensuring software quality and reliability during the development phase is crucial for the success of these AI-driven systems. JIT defect prediction leverages code metrics, commit histories, and machine learning models to predict whether a code change will introduce a defect [2].

Artificial intelligence-based methods have had a significant impact in providing software engineering solutions, such as defect prediction [2] and repair [3], code summarization [4], and clone detection [5]. Both machine learning and deep learning algorithms have been dealing with the software defect prediction task. The authors in [6] proposed a classification model that blends the ensemble learning method XGBoost with SMOTE-Tomek sampling. In [7], a machine learning approach employs multiple linear regression to predict the defect density of future releases of open-source software.

Similarly, natural language processing (NLP) models have been applied to the JIT-SDP domain, leveraging their ability to understand and generate code. While these models have shown promising results, particularly in analyzing syntax and semantics of code changes, they face challenges in capturing the full context of software development, which often involves multiple data modalities such as commit messages, code changes, and developer activity. The rapid development of coding foundation models has made this area a significant field of research. Early breakthroughs like Codex [8] and AlphaCode [9] demonstrated impressive capabilities and were quickly commercialized through tools like GitHub Copilot. However, these models still struggle with handling the diverse and complex data involved in JIT-SDP tasks. Continuous efforts are being made to improve their performance, especially in terms of generalization and efficiency. For instance, GraphCodeBERT [8] enhances standard token-based models by incorporating structural code information, such as the abstract syntax tree (AST), during the pre-training phase, which improves its understanding of the hierarchical nature of code. In contrast, our proposed approach leverages multimodal learning to address the limitations of existing models by fusing multiple data types, including textual message, categorical, and contextual information, to enhance defect prediction accuracy. Multimodal learning refers to the integration of diverse data types, where each modality contributes unique information to improve the model's decision-making capabilities. In summary, this work makes the following contributions:

- Multimodal JIT-SDP model using pre-trained transformer architectures to deal with the multimodal data of the SDP datasets like commit message, code metrics etc.
- The proposed framework also leverages task-specific language models *CodeBERT* and *GraphCodeBERT* which are pre-trained on the CodeSearchNet dataset, to check the reliability of multimodal learning.
- Multimodal transformers consist of multiple ways to combine feature methods using the combining module.
- Three different autonomous driving defect datasets are used to check the scalability and generalizability of the proposed model.
- In the experiments conducted, we have shown that the multimodal learning-based models provide comparable results to the state-of-the-art models and outperform the traditional SDP models.

## II. RELATED WORK

### A. Just-in-time Defect Prediction using Transformers

Software defect prediction is a critical task in software engineering that aims to identify potential defects in software systems before they cause significant issues. Pretrained models, which have been trained on large datasets and fine-tuned for specific tasks, can provide valuable insights and predictions for software defect detection.

Just-in-Time (JIT) defect prediction aims to identify defects as soon as they are introduced, enhancing the software development process's efficiency and reliability. Just-in-time defect prediction is a more precise method of defect prediction than file-level prediction because it operates at the commit level. In general, commits are smaller than files. As a result, less code needs to be analyzed to detect the flaw. To optimize the time and effort while inspecting the code for errors, developers can also publish the modified code to the repository and concurrently verify if the commit is prone to errors or not [11]. Recent work also raises concerns that a lack of explainability of software analytics often hinders the adoption of software analytics in practice. Therefore, in [10], the authors propose PyExplainer, a local rule-based model-agnostic technique for explaining the predictions of JIT defect prediction models.

Pre-trained language models like deep bidirectional transformers namely bidirectional encoder representations (BERT) functions as a trained encoder that combines natural language (NL) and programming language (PL) words into a single vector representing their semantic meaning [9]. Without requiring significant task-specific architecture changes, the pre-trained BERT model can be improved with just one extra output layer to produce state-of-the-art models for a variety of tasks, including question-answering and language inference.

This is an example of an autonomously learnt feature. Pretrained task-specific models such as *CodeBERT*, *GraphCodeBERT*, and *UniXCoder* trained on the bimodal data *CodeSearch* [17], have been applied for several software engineering tasks, such as automated code repair, code summarization, and defect prediction. In the case of the JIT defect prediction task, fine-tuned CodeBERT achieved comparable performance with the state-of-the-art approach (CC2Vec) on the OpenStack dataset but still lacks generalizability among the different datasets. Table I illustrates the recent research based on pre-trained transformer models for software engineering-related tasks like defect prediction, bug detection, automatic code repair, etc. This table describes each method with its features and limitations.

### B. Multimodal Learning using Transformers

In the last decade, transformers have changed the paradigm of artificial intelligence, especially in natural language processing by outperforming all other deep learning and machine learning models. Transformers are good at handling text features as well as other features from another modality. In case of multimodal learning, transformers specifically the self-attention layers, can be trained on a modality with enormous data (a natural language corpus) to find feature representations that work for any sequence of data, allowing for downstream transfer to other modalities. We specifically aim to explore the generalization capabilities of pre-trained language models to additional modalities with sequential structure [27].

Multimodal learning aims to enable users to instantaneously modify cutting-edge transformer models for scenarios involving text and tabular data, which are frequently found in real-world datasets depicted in Fig. 1. While treating the additional image modality as additional tokens to the input, models like ViLBERT [16] and VLBERT [17] are essentially the same as the original BERT model for pre-trained transformers on images and text. Pre-training on multimodal text and image data is necessary for these models.

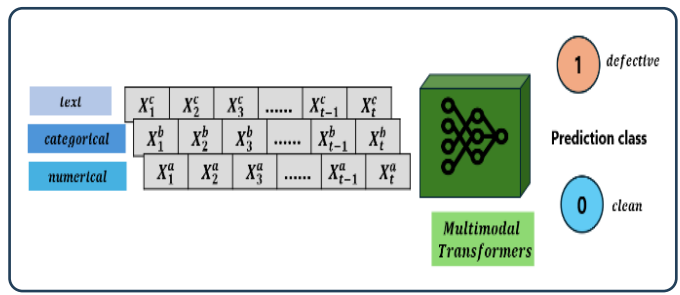

Figure 1. Multimodal learning using multiple data modalities of software defect prediction dataset

TABLE I. REVIEW OF TRANSFORMER-BASED MODELS FOR SOFTWARE ENGINEERING.

| Model | Description | Key Features | Limitations |
|---|---|---|---|
| GraphCodeBERT [12] | Transformer model leveraging code structure and relationships | Fine-tune for code defect prediction and understanding. Use in tasks requiring deep code structure understanding | Requires significant computational resources for training and inference. Complex model architecture may pose challenges in implementation and deployment. |
| CodeBERT [14] | Transformer-based model pretrained on source code and comments | Fine-tune on labeled defect datasets. - Integrate with development environments for real-time predictions | Requires large amounts of training data. Computational resource-intensive |
| UniXCoder [15] | Unified cross-modal pre-trained model for code understanding | Use for tasks combining code and natural language. Fine-tune for specific tasks like defect prediction | Resource-intensive, and complex to integrate into existing pipelines |
| DeepJit [25] | Uses deep learning for JIT defect prediction, focusing on real-time defect identification. Integration of deep learning models into development workflows. | Utilizes deep learning for real-time defect identification. Potential for integration into software development pipelines for continuous improvement. | - Requires large amounts of labeled data for effective training. Performance heavily reliant on quality and diversity of training data. |

Transformers have also been adapted for text, visual, and audio formats where alignment with the natural world is possible. In [19] multimodal transformers for unaligned multimodal language sequences (MulT) deploy three transformer architectures, each for one modality to capture the interactions with the other two modalities in a self-attentive manner. The information from the three Transformers are aggregated through late-fusion. MulT uses co-attention between pairs of modalities in addition to temporal convolutions to make input tokens aware of their temporal neighbours. Rahman et al. [20] use a gating technique to introduce cross-modal attention at specific Transformer layers. The above architectures have shown the limitation in offering a much more representative capability with the fixed interactive direction and restricted number of involved modalities, e.g., only up to two modalities. This indeed overlooks the complex and essential global interactions among more modalities, which results in information loss and the deterioration of prediction. Essentially, their model attempts to sequentially stack distinct two-way cross-modality attention blocks into the hierarchical one for the multimodal learning task, i.e., (text → visual) → acoustic. Intuitively, this sequential one-to-one procedure $\{Modality_i\} \rightarrow \{Modality_j\}$ would lead to a significant increase of cost in computation and memory. To overcome the above underlying research issues, we propose the multimodal transformer, that extends the standard transformer framework to analyze multiple modalities simultaneously

Multimodal Transformers suffer from two major efficiency issues: (1) Due to the large model parameter capacity makes them dependent on huge-scale training datasets. (2) They are limited by the time and memory complexities that grow quadratically with the input sequence length, which are caused by self-attention. In multimodal contexts, calculation explosion will become worse due to jointly high-dimensional representations. These two bottlenecks are interdependent and should be considered together. To improve the training and/or inferring efficiency for multimodal transformers, recent efforts have attempted to find various solutions, to use fewer training data and/or parameters which can gradually deal with the memory and time constraints.

III. PROPOSED APPROACH

This section describes the proposed method in detail. Fig. 2 illustrates the overall framework of our proposed approach. The framework is divided into the following four steps.

i. First, the defect data were collected by mining ADS data from the GitHub repositories using the PyDriller wrapper.
ii. Second, using the MA-SZZ [24] algorithm dataset labelling has been done.
iii. Third, data preprocessing to clean the data by

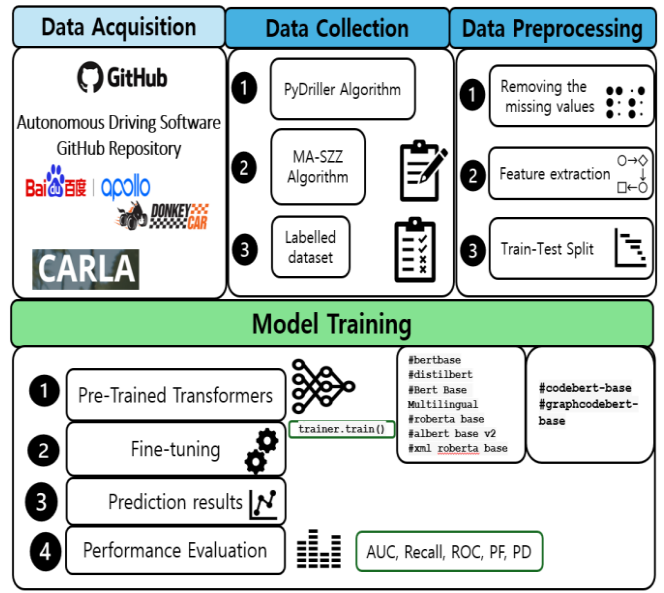

Figure 2. Multimodal Transformers for JIT-SDP Framework

removing the anomalies and missing values. Feature extraction and dataset split are also performed in this step.

iv. Finally, multiple MM-Trans are adapted to learn a multimodal representation of the ADS software defect dataset.

*A. Data Acquisition and Data Collection*

Data acquisition from the GitHub repositories by obtaining high-quality data which contains reliable information about the defects, code changes, and fixes in a particular software system is a tedious task. In this study, PyDriller wrapper (a Python framework) is employed to ease the extraction of the information from the GitHub repositories. On GitHub, we compile statistics about open-source self-driving software. When searching GitHub repositories, we enter the keywords self-driving and autonomous driving. After that, the three projects with the most stars are selected. The popularity of a repository on GitHub is indicated by its star rating (the higher the star rating, the more popular the repository). Table V shows the description of the dataset used in this study. PyDriller is simple and robust for mining software repositories (MSR) by considering only important features. The objective is to collect data from AI-enabled systems like autonomous driving repositories. These projects are prone to defects which can lead to heavy loss of life and effect the reliability of the autonomous cars. Three open-source autonomous driving software defect datasets are taken into consideration for checking the scalability, robustness, and reliability of the proposed model. In GitHub, the data collected from the repository is source code, issues, and commit information.

TABLE II. CHANGE LEVEL FEATURES OF THE AUTONOMOUS DRIVING SYSTEM DATASET

| Dimension | Feature | Description |
|---|---|---|
| Diffusion | NS | The number of modified subsystems |
| | ND | The number of modified directories |
| | NF | The number of modified files |
| | Entropy | Distribution of modified code across each file |
| Size | LA | Lines of code added |
| | LD | Lines of code deleted |
| | LT | Lines of code in a file before the change |
| Purpose | FIX | Whether or not the change is a defect fix |
| History | NDEV | The number of developers that changed the modified files |
| | AGE | The average time interval between the last and current change |
| | NUC | The number of unique changes to the modified files |
| Experience | EXP | Developer experience |
| | REXP | Recent developer experience |
| | SEXP | Developer experience on a subsystem |

Afterwards, the MA-SZZ algorithm is applied to automatically identify faulty entities. The output of the algorithm is the final dataset labelled as to which class each instance belongs either buggy or non-buggy later represented by 1 or 0 after data cleaning. Data metrics in the existing open-source systems include product measures such as complexity, process measures including the number of lines added/deleted/updated, and developer capabilities, shown in Table II. This research project confirms whether software metrics are suitable for existing open-source AI-based systems, and further identifies and proposes new measures considering the characteristics of AI-based systems.

*B. Data Preprocessing*

Data preprocessing for a Just-in-Time (JIT) defect prediction dataset from autonomous driving software systems involves several steps. These steps ensure that the data is clean, consistent, and ready for model training. Below are the typical steps you might take for preprocessing such a dataset. Firstly, the data cleaning is done to remove or impute missing values and filter out irrelevant commits. The second step is the feature extraction, like textual features which consists of commit messages and code comments, categorical features which is the metadata present in the datasets (like commit-id, code-change, code-fixes, etc.), and numerical features which have abundant features because it contains the code metrics like entropy, lines of code added/removed, and other complexity measures. The third step is the data transformation which includes tokenization using transformers like BERT or other variants like CodeBERT etc. for the embedding of text data, encoding of categorical data, and normalization or standardization of numerical data. The final step in data preprocessing is data splitting, which divides the dataset into training, validation, and test sets. In our method, the train, validation, and test split are 8:1:1.

*C. Multimodal Transformer for JIT-Defect Prediction (MMTrans-JIT)*

Just-in-time software defect prediction is a critical task in software engineering that aims to identify potential defects in code at the time of code changes. Traditional approaches to JIT-SDP primarily rely on features derived from the code itself or its associated metadata. This holistic perspective can potentially improve the accuracy and robustness of the proposed approach. Multimodal transformers are a class of models designed to handle and integrate data from multiple modalities. The architecture typically consists of separate encoders for each modality followed by a fusion mechanism that combines the encoded representations into a unified vector for prediction, how the end-to-end deep learning paradigm can be leveraged for simultaneous text, numerical, and categorical inputs by extending standard transformer networks into multimodal transformer networks that jointly operate on both text and tabular features [27-29]. Three multimodal network variants are used for each pre-trained transformer architecture, as depicted in Table III.

*1) General Purpose Pre-trained Transformer Module*

*a) BERT:* In the case of BERT, it consists of multiple layers of transformer encoders. Training and inference are made possible by the concurrent processing of the input text by each encoder layer. The feedforward neural networks in each layer are usually followed by a self-attention mechanism, which allows the model to capture hierarchical representations of the input text. It is also called the Masked Language Model (MLM), BERT randomly masks some of the words in a sentence and trains the model to predict the masked words based on the context provided by the surrounding words. BERT possess bidirectional context, unlike previous models like Word2Vec or GloVe, which generate fixed-size word embeddings based on the context of individual words, BERT captures the bidirectional context of words by considering the entire sentence. This allows BERT to understand nuances such as word sense disambiguation and syntactic/semantic relationships within a sentence.

*b) DistilBERT:* This model is a pre-trained model from the Hugging Face library. It is a smaller, faster, and cheaper version of BERT (Bidirectional Encoder Representations from Transformers) developed by Hugging Face. It retains 97% of BERT's performance while being 60% faster and 40% smaller. It is trained using knowledge distillation, where a smaller student model DistilBERT learns to mimic a larger teacher model BERT [32].

*2) Domain-specific Pre-trained Transformer Module*

*a) GraphCodeBERT:* GraphCodeBERT is a pre-trained model designed for source code understanding and generation tasks. It incorporates both the token sequences and the data flow within the code, effectively capturing syntactic and semantic information. This model enhances the performance of various code-related tasks like code completion, summarization, and translation.

*b) CodeBERT:* CodeBERT is a bimodal pre-trained model for programming languages, based on BERT. It is designed to understand the relationship between natural language and programming language, enabling tasks such as code search, code documentation generation, and code translation. It is trained on a large corpus of natural language and source code pairs.

*3) Fusion Layer/Combining module:*

This module combines the outputs from the modality-specific encoders. A model-independent combining module that receives as input, *x*, preprocessed categorical (*c*) and numerical (*n*) features, and text features generated from a transformer model. Three different combining modules used in this work are given in Table III, with their descriptions. The combining module then outputs a combined multimodal representation, *m*. Multimodal transformers incorporate cross-modal attention inside middle transformer layers. Three Methods include concatenation, attention mechanisms, or more sophisticated techniques like cross-attention [27].

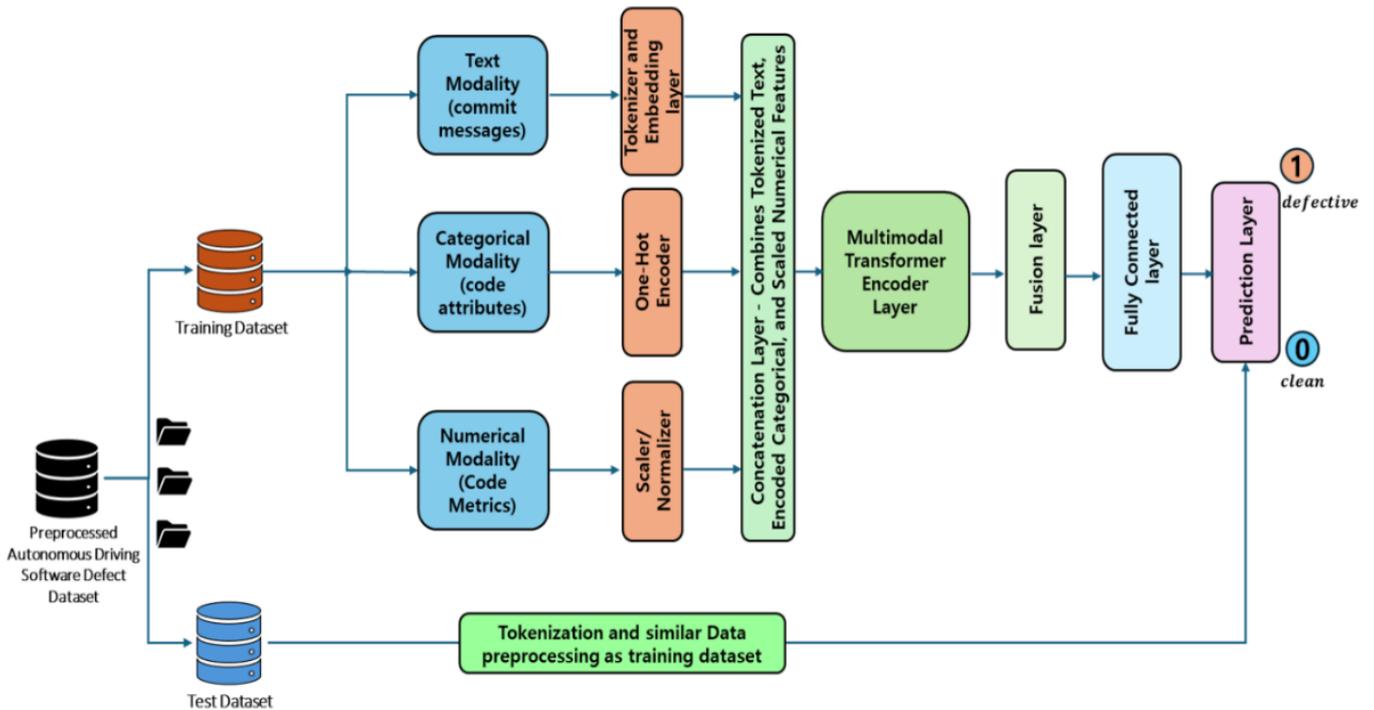

Figure 3. Implementation of the multimodal pre-trained transformers workflow for just-in-time software defect prediction

TABLE III. ILLUSTRATION OF THE COMBINING MODULE USED IN THE MM-TRANS FRAMEWORK

| Combine Feature Method | Description |
|---|---|
| Unimodal | Concatenate transformer output, numerical feats, and categorical feats all at once before the final classifier layer(s) |
| attention_on_cat_and_numerical_feats | Attention-based summation of transformer outputs, numerical feats, and categorical feats queried by transformer outputs before final classifier layer(s). |
| gating_on_cat_and_num_feats_then_sum | Gated summation of transformer outputs, numerical feats, and categorical feats before final classifier layer(s). Inspired by Integrating Multimodal Information in Large Pretrained Transformers [31] which performs the mechanism for each token. |

*4) Prediction Layer*: A fully connected feed-forward layer that takes the fused representation and outputs the probability of the code change containing a defect which categorizes it to either belong to defective class (1) or clean class (0).

## IV. EXPERIMENTAL SETUP

### A. Backbone transformer

The proposed supervised learning pipeline design is to determine the appropriate transformer backbone and fine-tuning strategy. The selection of the backbone depends on performance evaluated using the *unimodal* baseline. Fine-tune the pre-trained transformer models as sole predictors using only the text features in each dataset. This helps identify which model is better at handling the types of text in our multimodal datasets. Implementation details reveal that general-purpose pre-trained language models like BERT and DistilBERT perform better across the text columns because of their natural language processing ability. Therefore, we choose between the base version of BERT and DistilBERT. In the case of domain-specific language models like CodeBERT, and GraphCodeBERT two popular backbones possess the bimodal attribute to deal with the NL and PL because both are trained on the CodeSearch corpora.

### B. Algorithm

| Algorithm 1: Multimodal Learning for JIT Software Defect Prediction |
|---|
| **Input:** JIT-SDP dataset *D (X, y)* <br> **Output:** Evaluation metrics – *Accuracy, Recall, Precision, F1, AUC* <br> 1. **Load Datasets:** <br>   $Data_{train}$ ← LoadDataset ("Training Dataset") <br>   $Data_{test}$ ← LoadDataset ("Test Dataset`) <br> 2. **Preprocess the data for each modality** <br>   $T_{text}$←tokenize(commit_messages), <br>   $T_{cat}$←OneHotEncode(code_attributes), $T_{num}$←normalize(code_metrics) <br> 3. **Load Pre-trained Transformer Models:** <br>   *BERT, DistilBERT, CodeBERT,* and *GraphCodeBERT.* <br> 4. **Concatenate Features and Feed into Multimodal Transformer:** <br>   $concat_{feats}$ ← concatenate ($T_{text}, T_{cat}, T_{num}$) <br>   $Enc_{output}$← mTransEncLayer($concat_{feats}$) <br> 5. **Pass $Enc_{output}$ through Combine Module:** <br>   $fused_{feats}$ ← fusionLayer($Enc_{output}$) <br> 6. **Output Prediction:** <br>   $pred_{labels}$ ← predictionLayer(fully_connected_output) <br> 7. **Train the Model and Evaluate on Test Data:** <br>   Compute evaluation metrics ($Data_{test}$: accuracy, recall, precision, F1, AUC). |

### C. Evaluation Metrics

Five measures were used to evaluate the effectiveness of the classification: accuracy, recall, precision, F1 score, and AUC score. Accuracy is the most common metric in binary classification. However, F1 and AUC are also widely used in SDP tasks to deal with class imbalance. As can be seen in Table IV, the confusion matrix is counted by the actual labels and predicted labels.

TABLE IV. CONFUSION MATRIX

|  | **Predicted Negative** | **Predicted Positive** |
|---|---|---|
| **Actual Negative** | True Positive | False Positive |
| **Actual Positive** | False Negative | True Negative |

*1) Precision*

Precision is the ratio of true positive predictions to the total number of positive predictions (both true positives and false positives). It measures the accuracy of positive predictions.

$$Precision = \frac{TP}{TP + FP} \quad (1)$$

Precision is crucial when the cost of false positives is high.

*2) Recall (Sensitivity or Probability of Detection (PD))*

Recall is the ratio of true positive predictions to the total number of actual positives (true positives and false negatives). Recall is important when missing a positive case is costly. In SDP, high recall ensures that most of the actual defects are identified, which is critical for maintaining software quality and reliability.

$$Recall = \frac{TP}{TP + FN} \quad (2)$$

*3) F1 Score*

The F1 Score is the harmonic-mean of precision and recall. It provides a single metric that balances the trade-off between precision and recall. The F1 Score is useful when there is an uneven class distribution, providing a more balanced measure than precision or recall alone. In SDP, it helps to ensure that both false positives and false negatives are minimized.

$$F1 = \frac{2 * Precision * Recall}{Precision + Recall} \quad (4)$$

*4) PR-AUC (Precision-Recall Area Under the Curve)*

It measures the area under the Precision-Recall curve, which evaluates a model's ability to distinguish between positive (defective) and negative (non-defective) samples, especially in imbalanced datasets.

## V. THREATS TO VALIDITY

As with any empirical study, there are threats to validity that should be discussed. Below we discuss the external, internal, construct and conclusion validity of our study.

## A. External validity

Examining whether and to what extent the research findings can be generalized is the concern of external validity. The dataset employed in this study has a class imbalance issue. Therefore, to get rid of model overfitting, resampling is preferred. On the other hand, the GHPR (GitHub Pull Request) method can be considered for data collection to create a balanced dataset. However, the experiments conducted in this research can also be performed with a different data set.

## B. Internal validity

To assess the advantage of using the proposed model for predicting defective software modules, it needs to be compared with the recent prediction techniques based on language models. To improve the model's behaviour, the dataset needs to go through the class balancing techniques. However, SMOTE technique to balance the dataset can be used in comparison with the proposed model.

## VI. RESULTS AND DISCUSSIONS

In this study, multiple multimodal transformers for JIT-SDP have been experimented with. For the sake of simplicity, the multimodal transformer models are represented as *m-BERT, m-DistilBERT, m-CodeBERT,* and *m-GraphCodeBERT*. The experimental results demonstrate that our approach can effectively identify potential defects with high precision and recall. The integration of semantic information from commit messages significantly improved the prediction accuracy. Additionally, real-time feedback enabled developers to address issues promptly, reducing the overall defect density in the software. Lacking public benchmarks, academic research on ML for multimodal text/tabular data has not matched industry demand to derive practical value from such data. This paper provides evidence that generic best practices for such data remain unclear today, we simply evaluated a few basic strategies on our benchmark and found a single automated strategy that turns out to outperform the current deep learning techniques on the binary classification task for finding the defective and non-defective change level software defects involving diverse text/tabular data. This strategy uses a stack ensemble of tabular models trained on top of predictions from other tabular models and a multimodal transformer.

### RQ1: Which multimodal transformer model achieves the best performance among the experimented approaches?

Among the multimodal transformer architectures, the *m-DistilBERT* is comprised of the DistilBERT as the transformer with the multimodal combining module uses a gating mechanism on the numerical and categorical features and then performs the sum with the text features to produce the output. This model performed better than its variants in the DonkeyCar dataset in terms of accuracy and PR-AUC, which led us to select it as the best candidate for the JIT-SDP task.

### RQ2: Does the proposed approach outperform the existing machine learning models?

The six multimodal-based models show comparable or better performance when compared with the existing models machine-learning models in some evaluation metrics as shown in Figure 4. It depicts that the multimodal based models (BERT-base-uncased, DistilBERT, CodeBERT, and GraphCodeBERT) consistently outperform the non-multimodal based transformer models (CodeBERT, and GraphCodeBERT without combine/fusion module) across all three datasets, especially in F1-score, precision, and accuracy. GraphCodeBERT stands out in terms of balanced performance (high F1, precision, and recall) compared to the other models. The performance usually depends on the amount and quality of the data.

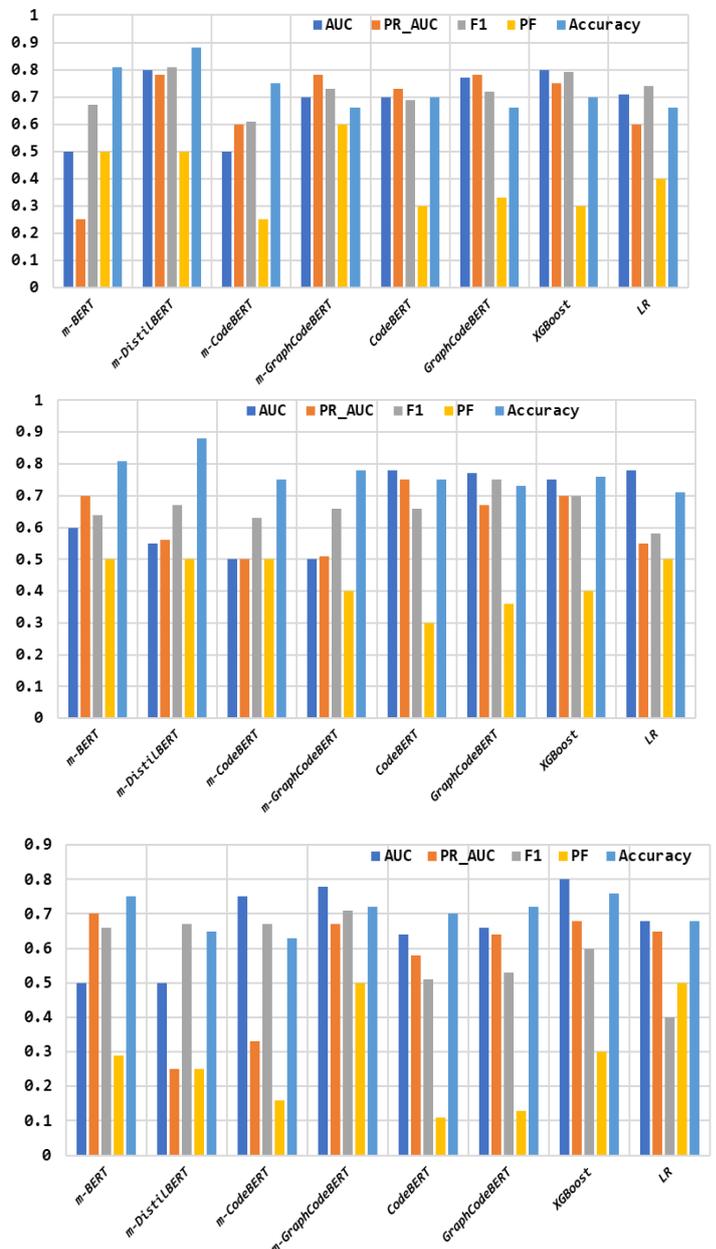

Figure 4. Performance chart for transformers against the baseline models (a. Carla dataset, b. multimodal DonkeyCar dataset, and c. Apollo dataset)

## VII. Conclusion

The integration of multimodal transformers for Just-In-Time Software Defect Prediction (JIT-SDP) on datasets from autonomous driving platforms such as Apollo, DonkeyCar and Carla has shown promising results. The multimodal-based models (BERT, DistilBERT, CodeBERT, GraphCodeBERT) show consistent improvement in accuracy (7.55% to 17.31%) over the models with no multimodal combined module in all datasets. While the performance on F1-Score varies. The multimodal models outperform the Apollo dataset (8.96% better F1), while the last four models show better results for the DonkeyCar and Carla datasets. The results indicate that while the first four models have better precision and are generally more balanced, the last four models, especially XGBoost and LR, excel in the recall, meaning they capture more positive cases but are prone to false positives. By leveraging the strengths of multimodal transformers, which effectively handle diverse data types including code snippets, textual descriptions, and numerical features, we have achieved performance metrics on par with state-of-the-art (SOTA) models like CodeBERT and GraphCodeBERT.

Future work can explore the application of small language models (SLMs) and large language models (LLMs) to further enhance JIT-SDP. By incorporating sLLMs, we can leverage vast amounts of unlabeled code and textual data to improve the model's understanding of code semantics and context. Furthermore, resampling techniques either under-sampling or employing SMOTE can bring better results.

## Acknowledgement

This research was supported by the Basic Science Research Program (NRF- 2022R1I1A3069233) through National Research Foundation of Korea(NRF) funded by the Ministry of Education (MOE), and the MSIT(Ministry of Science and ICT), Korea, under the ITRC(Information Technology Research Center) support program(IITP-2025-2020-0-01795) supervised by the IITP(Institute for Information & Communications Technology Planning & Evaluation) and the Nuclear Safety Research Program through the Korea Foundation Of Nuclear Safety (KoFONS) using the financial resource granted by the Nuclear Safety and Security Commission(NSSC) of the Republic of Korea. (No. 2105030)